# Minimization of THD in Nine Level Cascaded H-Bridge Inverter Using Artificial Neural Network

**Manoj Mathews**
Senior Authority Liaison Engineer,
Arcadis Consulting Middle East Limited,
manojmathews123@gmail.com

**B. Ramesh**
Lecturer, Department of EEE
NPA Centenary Polytechnic College, Kotagiri,
The Nilgiris Dist - 643217
rameshnpacpc@gmail.com

**Dr. T. Sreedhar**
Executive Director, Nice Panel Global Technologies,
Tirunelveli, Chennai.
sree822@gmail.com

*Abstract*—Multilevel inverter converts different level DC voltage to AC voltage. It has wide interest in power industry especially in high power applications. In power electronic equipments the major drawback is the harmonics. Several control strategies are available to reduce the harmonic content and the most widely used measure of Total Harmonic Distortion (THD). In this project the comparison has been made for the open loop and closed loop PI controller and neural network that predict the switching angle in order to reduce the harmonics. The mapping between Modulation Index and Switching angles are plotted for the forward neural network. After the prediction of switching angles the neural network topologies are executed for better result. This technique is applied for any type of multilevel inverter, Cascaded H-Bridge multilevel inverter is chosen. A nine level Cascaded H-Bridge multilevel inverter power circuit is simulated in MATLAB 8.3 simulink with sinusoidal PWM technique. The comparison results reveal that the THD is reduced to about 3% with neural network control compared to open loop control. The results are presented and analyzed.

## I. INTRODUCTION

POWER ELECTRONICS (Peter Cooper Hewitt, 1902) is applied to control and convert the electric power. Now-a-days it is also a subject of research in electronic and electrical the deals with design, control, convert, integrate and all other time varying electronic system appliances with fast dynamic response. A multilevel inverter (Baker and Bannister. 1975) is a power electronic device which could be preferred alternating voltage at the output using multiple lower level DC voltages as input. Mostly, a two-level inverter is used in order to generate the AC voltage from DC voltage [1], [3]. The main negative aspect of the two-level inverter is harmonic content in the output current this can be reduced by increasing level of the output voltage and current by series connecting multiple H-bridge inverter. The PWM inverters (1976, Bob Mammano) are used to control the output voltage and frequency simultaneously with the reduction of harmonic content in the load current [2], [18].

These significances are widely used in various industrial application such as Uninterrupted Power Supply, Variable Speed Drive and other power conversion system. In high power application, multilevel inverter is used for conversion of electrical power hence many high power appliances such as Variable speed drives and inverters for solar power generation, multilevel inverters are used [3]. Multilevel inverter converts different level DC voltages to AC voltage. Each DC voltage level adds a step to the AC voltage waveform, these DC voltages may or may not be equal to one another. There are three main types of multilevel inverter: diode clamped, flying capacitors and cascaded H-bridge. The cascaded h-bridge multilevel inverter topology has many constructional benefits so it is preferred in many conversional solid state electronic equipment. Keith Corzine (2002) suggested a general idea of cascading any number of multilevel H-bridge cells. It can be used in medium voltage electrical drives, electric vehicle and grid connection photovoltaic systems. The proposed inverter can reduce the harmonic content using sinusoidal PWM technique under the condition of equal DC voltage sources. The comparison is also made for open loop, closed loop PI and neural implementing Sinusoidal Pulse Width Modulation Technique (SPWM) technique. The control strategy involves the generation of PWM signal for the modulation index (m) between the range of 0.1 to 1. The closed loop control of cascaded multilevel inverter is also implemented by simulating conventional PI controller [4].

An image of a hybrid cascade converter topology with sequence-connected symmetrical and asymmetrical diode clamped H-bridge cells are researched by Nami et al [5]. A new H-bridge multilevel pulse width modulation converter topology based on a series banding together of a high-voltage diode-clamped inverter and a low-voltage approved inverter is approaching in this paper. A dc equal voltage spot for the dressed to the teeth cur and asymmetric consolidation is spotted to have a maximum abode of annual production voltage levels by preserving the neighboring switching vectors mid voltage levels. Hence, a 15-level hybrid converter cut back be attained by the whole of a minimum abode of art components. Voltage sharing converter to supply single-phase asymmetrical four-level diode clamped inverter with high power factor loads [6].

In this complimentary, a nifty single-inductor multi-output dc/dc converter is about to handle on something the dc-link voltages of a single-phase diode-clamped inverter asymmetrically to get ahead voltage action enhancement. The trek of the invented converter is explained and the dominating equations are developed. A behave strategy is approaching and explained in details. To vindicate the versatility of the proposed mishmash of the tacit dc-dc converter and the jagged four-level diode-clamped inverter (ADCI), simulations and experiments have been directed. Modeling and analysis of switching-ripple voltage on the DC link between a diode rectifier and a modular multilevel cascade inverter (MMCI) *Authors*:Hui Peng et al [7].,This paper describe the common dc-link voltage that flows between a three phase diode rectifier and a cascaded inverter bsed on double-star chopper cells (MMCI-DSCC) for a medium voltage drive which can be operated even there is no capacitor.

Finally, the switching angles are predicted that occurs on the dc-link. Therefore, a small sized dc passive filter consisting of series connection of a thin film capacitor and a damping resistor is achieved. Charge balance control of a low resolution symmetric subsystem for an asymmetric cascaded multilevel inverter [8], [17].

To achieve the possibility of charge balance control method application the topology of an asymmetric cascaded multilevel inverter has been developed.

In this paper, a new algorithm is implemented for the determination of dc voltage magnitudes at all levels. To equalize the losses value in switches related to each bridge in each unit, two conventional charge balance control methods have been developed to symmetrically charge the dc sources. A single phase multilevel inverter using switched series/parallel dc voltage sources [9].

A multilevel inverter with less number of switches are developed. It consists of an H-bridge inverter in which the switches are connected in series and in parallel to the dc voltage source. This proposed system same output is attained for circuit with reduced switches and also switching losses can also be minimized.

The total harmonic distortion of the output waveform is also reduced. The number of triggering circuits are reduced, which leads to the reduction of the size and power consumption in the driving circuits. The inverter is driven by the hybrid modulation method. The circuit configurations are simulated using MATLAB/ Simulink.

## II. TECHNIQUES TO REDUCE HARMONICS

The solid state electronic equipments provide voltage and current harmonics, where the output current with lower order harmonic is the serious factor which has to be eliminated [10]. These harmonics affect the equipment and the particulars that are connected in series and parallel to the inverter.

There are two types of harmonics that is, the even order harmonics and odd order harmonics. In which the even order harmonics are eliminated by using filters and the odd order harmonics can be eliminated by various control strategies and techniques. The harmonic content and harmonic distortion is analyzed by the following techniques.

### A. PWM Control

The performance of PWM control strategies is based on the following parameters:
a. Total harmonic distortion (THD) of the output voltage and current in the inverter,
b. Switching losses that occurs due to the presence of ON/OFF power semiconductor devices within the inverter,
c. Peak-to-peak ripples that occurs in the load current, and
d. Maximum output voltage of an inverter for given DC input voltages.

Therefore, depends on the harmonic content the PWM Control technique was chosen in the inverter output voltage. As mentioned above PWM control technique has many advantages compared to the other method of control. Sinusoidal PWM inverters play a key role to control amplitude, frequency and harmonics contents in the output voltage and current. The SPWM aims to produce sinusoidal inverter output voltage without low-order harmonics.

Sinusoidal pulse width modulation is one of the primitive techniques, which is used to reduce harmonics presented in the quasi-square wave. There is an important parameter in the modulation techniques, i.e., the ratio:

$$M = A_r/A_c \text{ known as modulation index,}$$

where $A_r$ is reference signal amplitude; $A_c$ is carrier signal amplitude.

*B. Harmonic Elimination Switching Angles (HESA)*

It is assumed that the HESA is quarter-wave symmetric. For a cascaded H-bridge inverter the Fourier series of the quarter-wave symmetric is written as follows:

$$V(wt) = \sum_{n=1}^{\infty} \frac{4V_{dc}}{n\pi} \left[ \sum_{k=1}^{s} \cos(n\theta_k) \right] \sin(nwt)$$

where the optimized switching angles must satisfy the following condition:

$$\theta_1 < \theta_2 < ... < \theta_s < \frac{\pi}{2}$$

The amplitude of all odd order harmonics including fundamental one, are given by:

$$h(n) = \frac{4V_{dc}}{n\pi} \sum_{k=1}^{s} \cos(n\theta_k)$$

The switching angles given to each switch are adjusted for reducing the output voltage THD. If peak value of the output voltage V1 has to be controlled, eliminate the third and fifth order harmonics. Now the modulation index is given by:

$$M = \frac{\pi V_1}{4V_{dc}}$$

The resulting lower order harmonic equations are:

$$\frac{4V_{dc}}{\pi}\left[\cos(\theta_1) + \cos(\theta_2) + \cos(\theta_3) + \cos(\theta_4)\right] = V_1$$
$$\left[\cos(5\theta_1) + \cos(5\theta_2) + \cos(5\theta_3) + \cos(5\theta_4)\right] = 0$$
$$\left[\cos(7\theta_1) + \cos(7\theta_2) + \cos(7\theta_3) + \cos(7\theta_4)\right] = 0$$
$$\left[\cos(\theta_1) + \cos(\theta_2) + \cos(\theta_3) + \cos(\theta_4)\right] = M$$

Harmonic elimination switching angles of 5th and 7th order for nine level cascaded multilevel inverter are solved by using Newton Raphson method [11].

*C. Neural Network Topology*

Neural networks are non-linear statistical data modeling tools. An artificial neural network (ANN), is also called as "neural network" (NN), it is a computational model or mathematical solution retrieval that is used to simulate the structure and functional aspects of biological neural networks. A neural network is an interconnected group of nodes, akin to the vast network of neurons in the human brain. It is an adaptive system. For computation it consists of an interconnected group of artificial neurons and processes information using a connectionist approach. In most cases, during learning phase it changes its structure based on internal or external information that flows through the computational network. They can also be able to compute complex relationships between inputs and outputs or to find patterns in different set of data [12].

1. Structure of Feed Forward Network

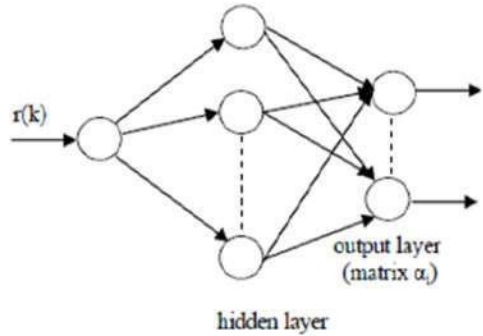

*Figure 1: Feed Forward Neural Networks*

The feed forward neural network as shown in Figure 1, the first and simplest computational method of neural network topology is the feed forward neural networks [13].

2. Training of Neural Network Using MATLAB Coding

A set of non-linear equations are solved to determine the switching angles of the Selective Harmonic Elimination Pulse Width Modulation. In case if two possible solutions of switching angles $\alpha_i$ are obtained then one among it is the Total Harmonic Distortion (THD). The accurate angles are therefore selected for the suppression of lower order harmonics. In many studies it has been proved that this training method using ANN has gained tremendous popularity in the angle prediction and reduction of THD compared to other alternative method.

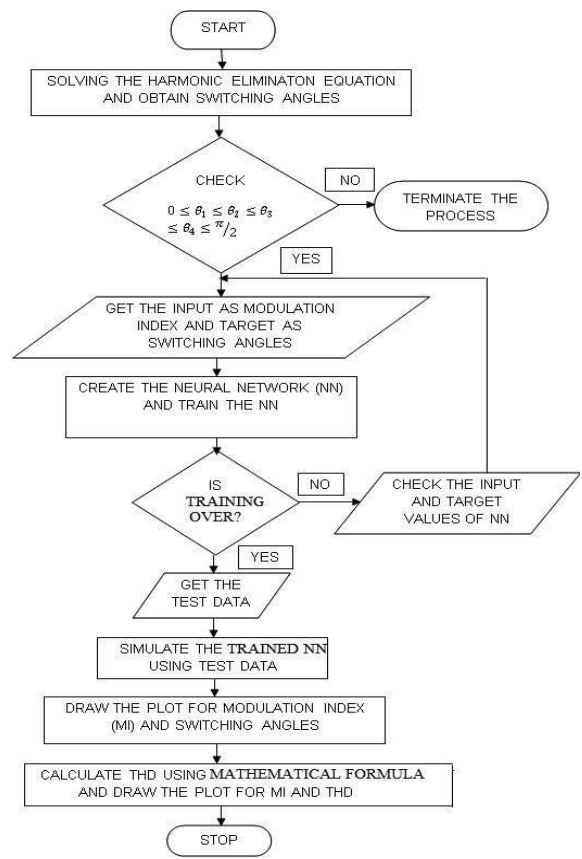

*Figure 2: Flow chart of artificial neural network technique*

In the supervised training phase, the correct class for each record is known and simultaneously the output nodes correct values are assigned to be "1" for each correct class with corresponding nodes and for other nodes it is declared as "0". In practical system it is better to use values of 0.9 and 0.1, respectively. So, it is possible to compare calculated network valued for each output nodes of these "correct" class and by using the delta rule an error term for each node is calculated. Now the error terms are used to vary the weights of in the hidden layers so that the next processing attempt will generate closer to the "correct" values. The flow chart representation of neural network implementation was shown in Fig. 2.

Certainly, without proper understanding of the process under investigation, artificial neural network can plot the relationship between and output data. This graphing is attained by the adjustments of their internal parameters called weights from data. This method of action is called the learning or the training process. Generalization capabilities are also set limits to their interest, i.e., heir exemption to put estimated responses to inputs that were not seen completely during the training period. Therefore, the ANN is applied in many complex relationship and processes that makes them highly attractive for various kinds of recent problems.

**TABLE I : THD VALUES FOR VARIOUS MODULATION INDEX AND THEIR SWITCHING ANGLES**

| MI | Switching angles in degree | | | | THD in % |
|---|---|---|---|---|---|
| | Theta 1 | Theta 2 | Theta 3 | Theta 4 | |
| 0.8 | 21.73 | 35.24 | 48.36 | 55.12 | 4.1 |
| 8.81 | 19.45 | 30.26 | 42.34 | 49.89 | 4.02 |
| 0.82 | 17.62 | 28.33 | 39.64 | 46.77 | 3.95 |
| 0.83 | 15.78 | 26.12 | 37.65 | 45.88 | 3.83 |
| 0.84 | 15.23 | 25.93 | 36.87 | 44.79 | 3.78 |
| 0.85 | 14.99 | 25.86 | 36.45 | 44.31 | 3.61 |
| 0.86 | 14.67 | 25.75 | 35.90 | 43.82 | 3.53 |
| 0.87 | 14.59 | 25.66 | 35.76 | 43.73 | 3.46 |
| 0.88 | 14.35 | 25.53 | 35.61 | 43.69 | 3.33 |
| 0.89 | 14.19 | 25.39 | 35.40 | 43.51 | 3.27 |
| 0.9 | 14.01 | 25.18 | 35.29 | 42.38 | 3.12 |
| 0.91 | 13.99 | 24.91 | 35.00 | 42.46 | 2.99 |

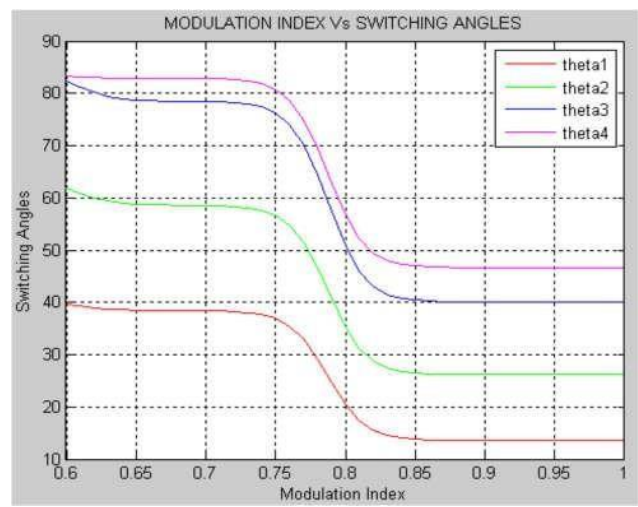

*Figure 3: Plot for modulation index vs corresponding switching angles.*

The THD values for various modulation index and their switching angles are shown in the Table I. From that it has been deduced that the increasing of Modulation Index results in reduction of THD values.

The Fig. 4 shows the response between modulation Index and THD values that are calculated for their corresponding switching angles. For about 1000 epochs the neural networks are trained foe all given set of inputs and the expected targets (desired switching angles calculated by solving harmonic elimination equation).

### III. CASCADED MULTI-LEVEL INVERTER

For satisfying the greater demand of medium voltage high power application cascaded multilevel inverter is used. It consists of multiple units of single-phase H-bridge inverter. To achieve medium voltage operation and low harmonic distortion the H-bridge cells are connected in cascade in their ac side. In requires a number of isolated dc input voltages which feeds each H-bridge power cells.

Thus, a cascaded multi-level inverter consists of a cascade or series connected H-bride inverter units with separate dc voltage source as shown in Fig 5 [14]. Each single-phase inverter consists of four switches S1, S2, S3, and S4 connected with dc input voltage that converts into ac output voltage.

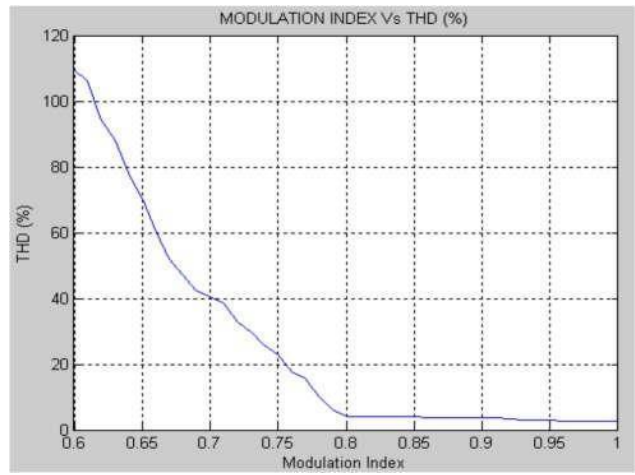

*Figure 4: Plot for modulation index and THD in percentage*

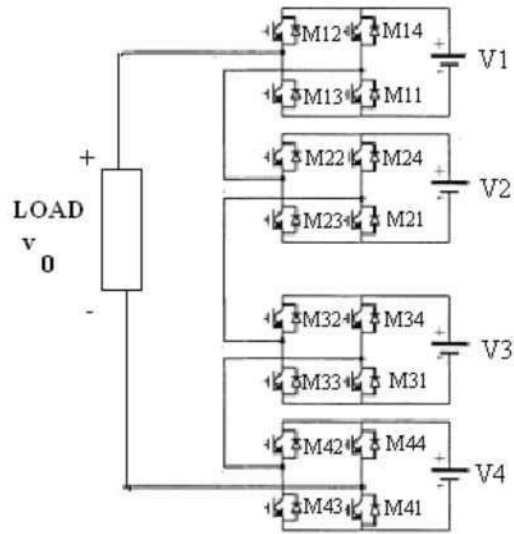

*Figure 5: Structure of a single phase cascaded nine level inverter*

#### A. Proposed System

The series connection of single phase full bridge inverter is the cascaded H-bridge multilevel inverter. The general operating principle of the multilevel inverter is to convert different level of dc voltages to ac voltage, which may also be obtained from batteries, fuel cells, or solar cells. In three types of multi-level inverter, cascaded H-bridge inverter type is considered for this work. In this topology, for a nine-level inverter four single phase H-bridge inverter are connected in series. To determine the number of bridges required for an "m" level inverter is ((m-1)/2).

For the DC voltage source magnitude and fundamental frequency the switches in the multilevel inverter are switched for its switching angle. In the cascaded multilevel inverter all the voltage sources separated from one another. Therefore, for nine level inverter four DC sources are needed. Because of its better switching utilization the switching stress can be reduced. In the proposed system for four H-bridges of the multi-level inverter identical DC source voltages are used. The structure of single phase cascaded nine level inverter is shown in figure 5. Each bridge structure consists of four Metal Oxide Semiconductor Field Effect Transistors (MOSFET).

Each single-phase H-bridge inverter level can produce three different output voltages +Vdc, 0 and –Vdc. The number of output phase voltage levels in a cascaded multi-level inverter can be predicted using (S*2) +1, where S is the number of dc sources [2], [15].

The voltage waveform of a nine-level cascaded inverter is shown in Fig. 6. The maximum output voltage is given by V0 = V1+V2+V3+V4. Each bridge unit generates a phase shifting its position and negative phase-leg-switching timings. It should be noted that each switching device always conducts for 180° (or half-cycle). This switching method makes the switching device current equal. The multilevel inverter results in an output voltage that is near sinusoidal with enough levels and an appropriate switching algorithm.

*B. Method of Working*

The output phase voltage of the cascaded-nine level inverter is shown in Fig. 6. The steps to generate the nine level voltages are as follows:
1. Switches M11, M12, M22, M32 and M42 are turned ON for an output phase voltage level V0 = V1,
2. Switches as mentioned in step 1 and M21 are turned ON for an output phase voltage level V0 = V1+V2,
3. Switches as mentioned in step 2 and M31 are turned ON for an output phase voltage level V0 = V1+V2+V3,
4. Switches as a mentioned in step 3 and M4 are turned ON for an output phase voltage level V0 = V1+V2+V3+V4.

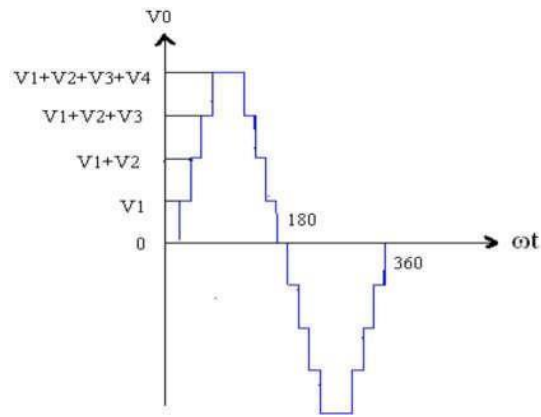

*Figure 6: Nine level inverter waveform*

## IV. SIMULATION RESULTS

The Simulink model of the proposed nine level cascaded multilevel inverter systems for SPWM techniques with open loop, closed loop PI and neural implementation are described by following simulation diagrams.

The simulation was done for a cascaded nine level inverter with sinusoidal PWM is described in Fig. 7. To obtain the output voltage with reduced THD, sinusoidal pulse width modulation technique is used.

*A. Sinusoidal Pulse Width Modulation-Open Loop*

A Sinusoidal Pulse Width Modulation technique is also known as the triangulation, sub oscillation, subs harmonic method which is used widely in industrial applications.

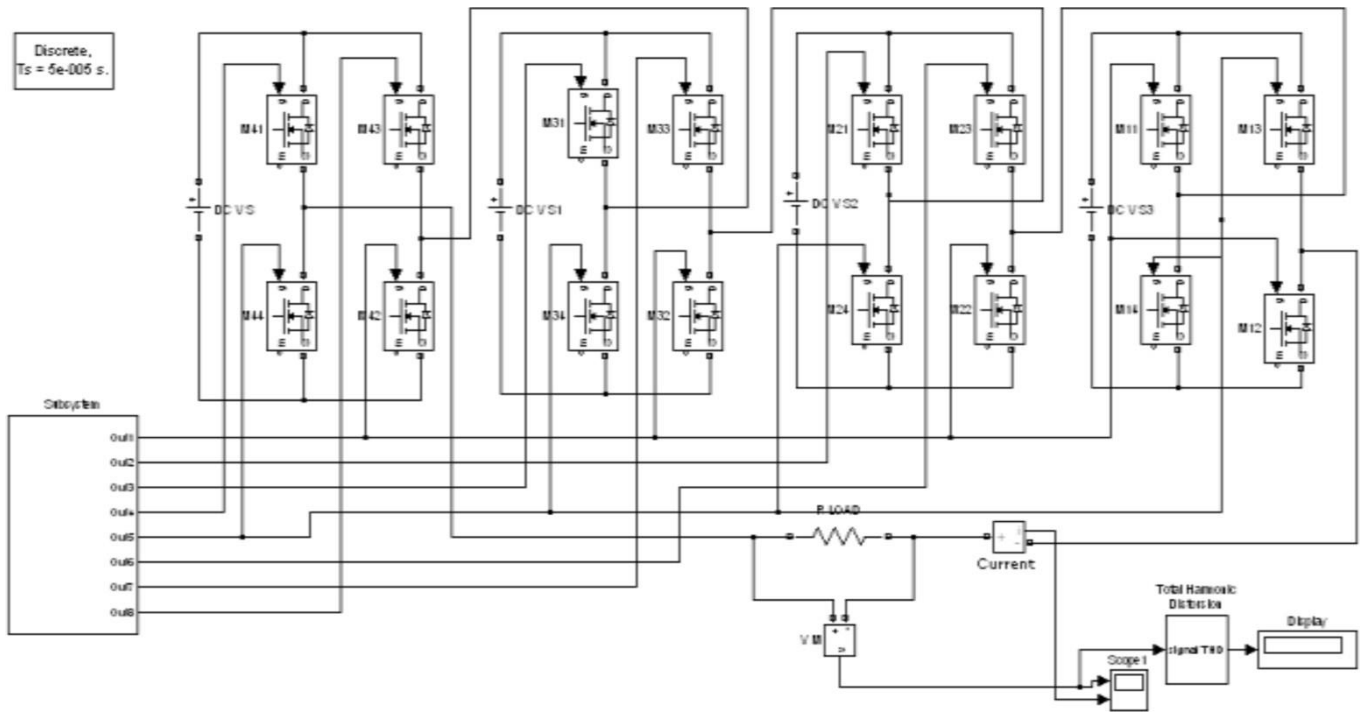

*Figure7: Simulink model for cascaded nine level inverter using sinusidal pulse width modulaiton technique*

In this technique a high frequency triangular carrier wave is compared with the sinusoidal reference wave the comparison result determines the switching angles for the reduction of switching losses to provide better dynamic response.

The simulated output voltage and current is shown in Fig. 8 and the frequency spectrum of output voltage is shown in Fig. 9.

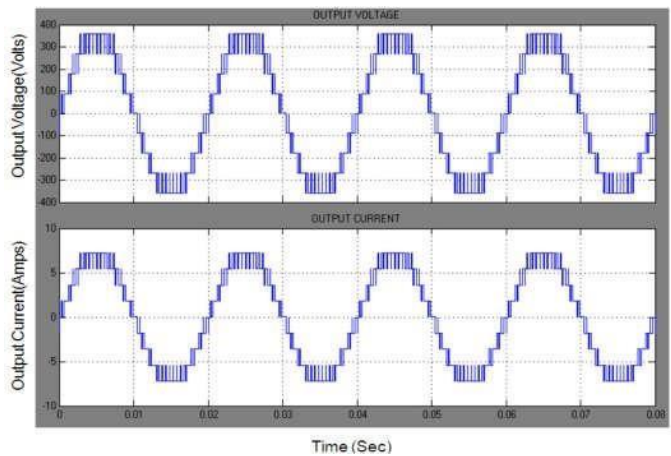

*Figure 8: Output voltage and current waveform for sinusoidal PWM*

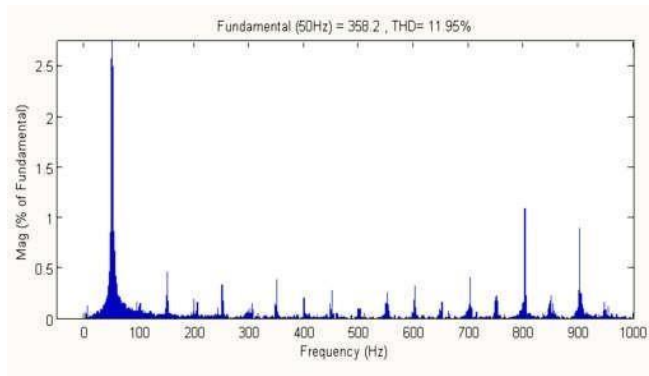

*Figure 9: Frequency Spectrum of the output voltage in sinusoidal PWM*

*B. Closed Loop- PI*

In closed loop, conventional PI controller is simulated as feedback system to control the inverter performance.

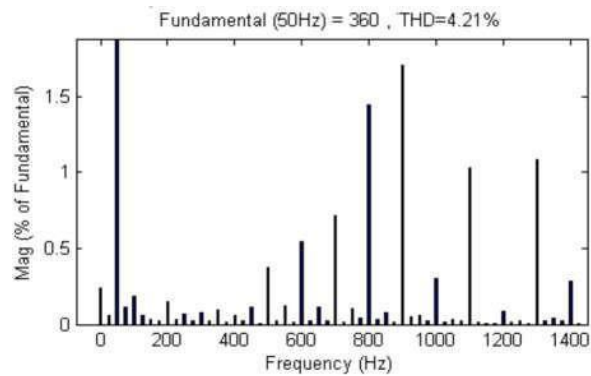

*Figure 10: Frequency spectrum of output voltage (PI controller)*

The gate signal is generated by the error signal retrieved for each comparison between the output voltage quasi-square wave signal and the reference sine wave. For various values of "m" the THD values are measured. Hence for modulation index m=0.9, the THD is 4.21%.

*C. Proposed Scheme for Nine Level Cascaded Inverter with Neural Network*

In view to confirm the validity of results obtained by harmonic elimination technique with neural network. From switching angles the desired level of carrier signal levels are obtained. The triggering pulse of each switch is generated by comparing the carrier signal and the reference signal. This pulse will make the switches to turn ON and hence the cascaded multilevel inverter will be operated.

**TABLE II : SPECIFICATION OF THE PROPOSED TOPOLOGY**

| | |
|---|---|
| DC voltage 1 | 10V |
| DC voltage 2 | 10V |
| DC voltage 3 | 10V |
| DC voltage 4 | 10V |
| R load | 100Ω |
| Switching Frequency | 5 kHz |
| Peak Output Voltage | 40V |
| No of level | 9 |

The PI controller based closed loop system is shown in Fig. 12. The output voltage is having nine level calibers of dc voltage. The output voltage and current are showing in Figs. 13-15 [16].

The multicarrier pulse generation is done by comparing the reference with the eight different carrier signals with 5 kHz. The output voltage from the proposed network topology is showing the best results in terms of minimum harmonic distortion levels. The total harmonic distortion with the help of FFT analysis is shown in Fig. 15.

## V. CONCLUSION

The reduced THD of the cascaded H-bridge nine level inverter by open loop, closed loop PI controller and neural network are analyzed, and comparison result are brought out. The comparison results reveal that the proposed artificial neural network minimized the lower order harmonics progressively. The neural network topology is operated based on the learning and approximating of the relationship between the modulation index and the switching angles with a feed forward network. The resulting neural network implementation for the elimination of harmonics has the advantages of very few computational cost and high-performance accuracy. Reduced total harmonic distortion makes the inverter system efficient in conversion and its power conversion applications.

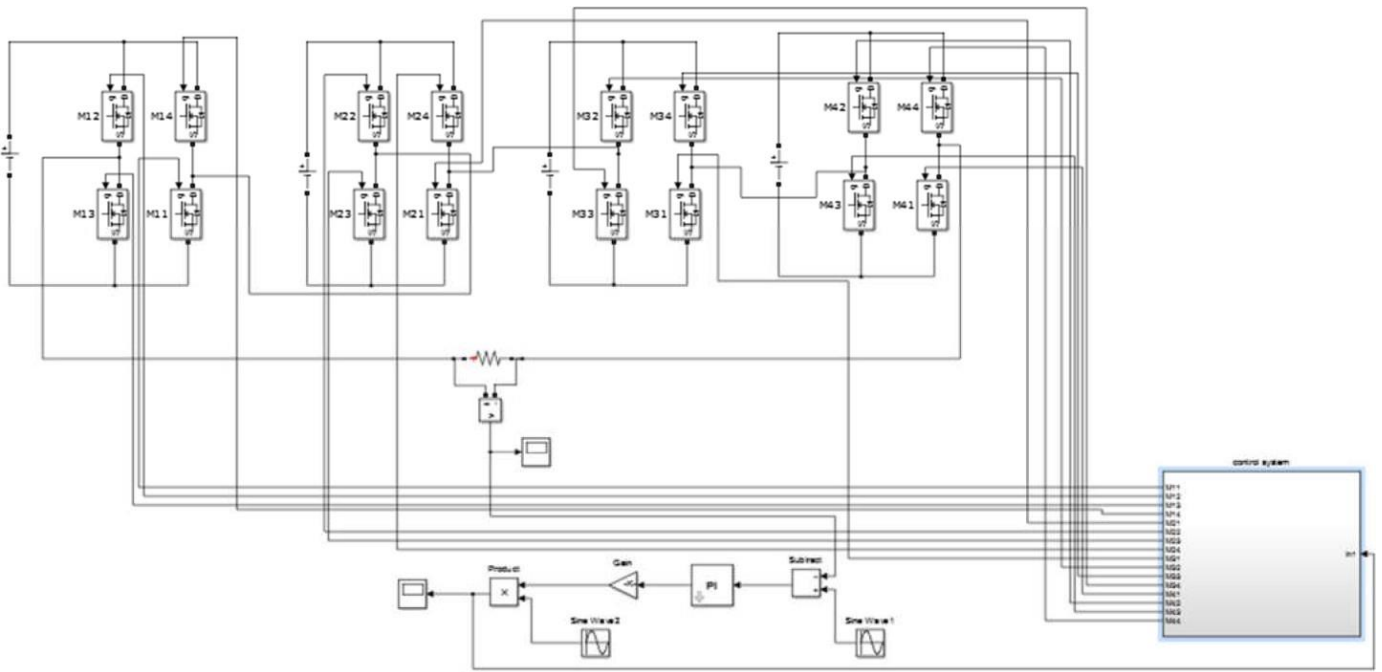

*Figure 11: Over all Simulink model for proposed system*

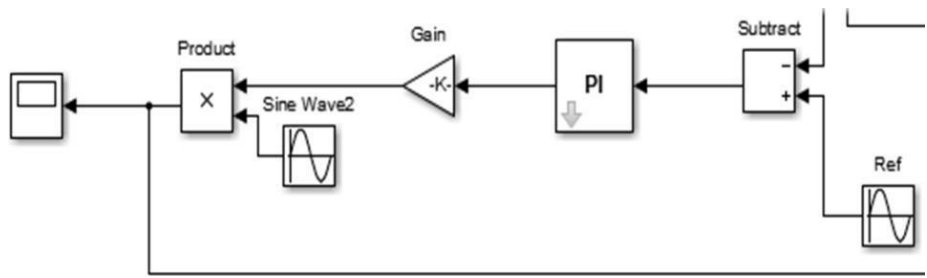

*Figure 12: Simulink model for PI based closed loop control*

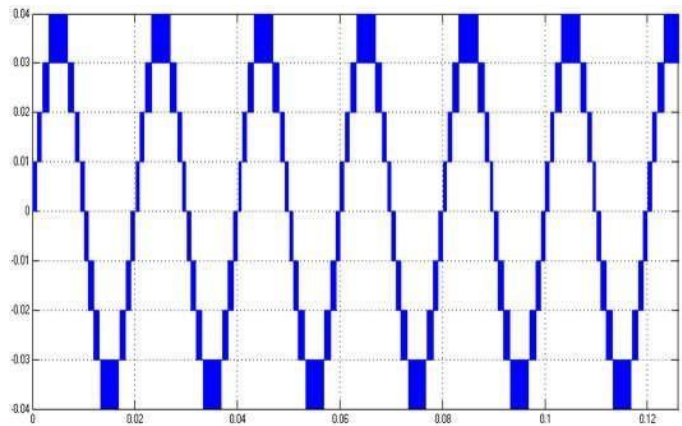

*Figure 13: Output current with nine level*

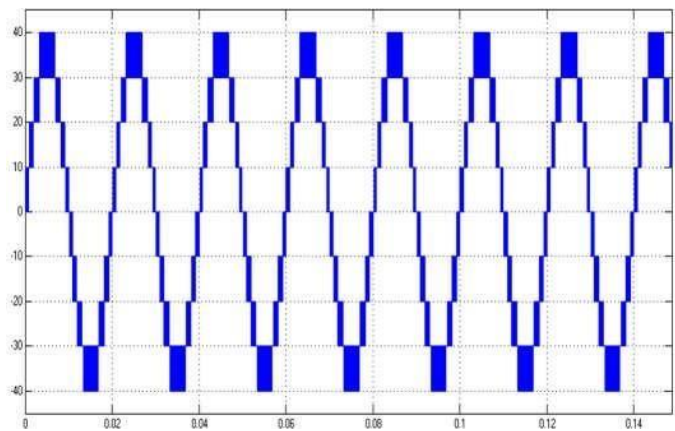

*Figure 14: Output voltage with nine level*

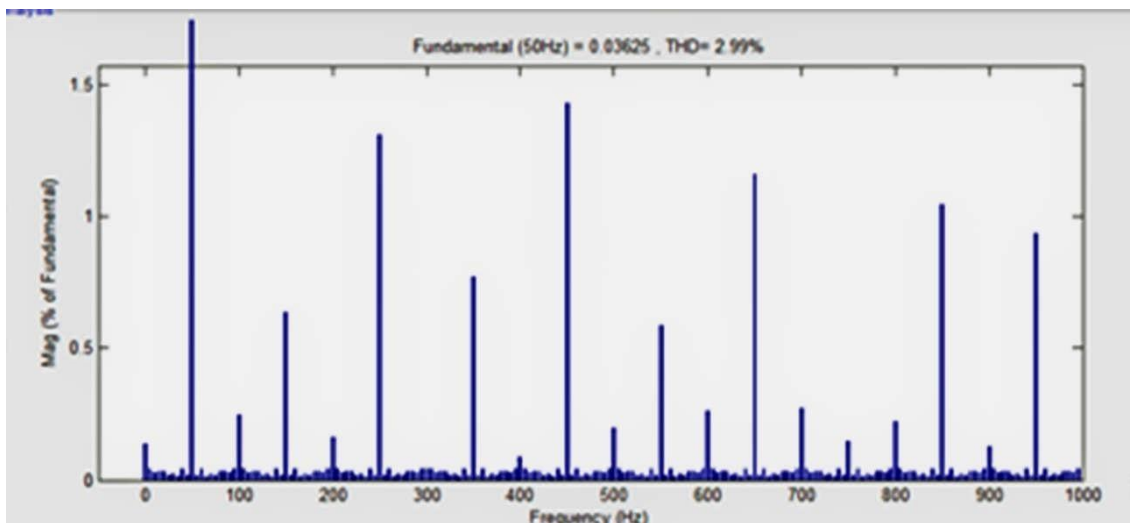

*Figure 15: THD in output current proposed topology*

**REFERENCES**

[1] Akira Nabae, I. Takahashi and H. Akagi, "A New Neutral-Point-Clamped PWM Inverter" IEEE Trans. Ind. Applicat. , vol 1A-17 , no. 5 , sep/oct. 1981 , pp. 518 – 523.